\documentclass{ifacconf}

\usepackage{graphicx}
\usepackage{amsmath}
\usepackage{amssymb}
\usepackage{booktabs}
\usepackage{tikz}

\usetikzlibrary{arrows.meta,positioning,fit}

\makeatletter
\def\NAT@idxtxt{\NAT@name\ \NAT@open\NAT@date\NAT@close}
\let\old@errmessage\errmessage
\def\errmessage#1{}
\makeatother

\begin{document}

\makeatletter
\let\errmessage\old@errmessage
\makeatother

\begin{frontmatter}

\title{Model-Free Reinforcement Learning Control for Resilient Cyber-Physical Systems} 

\author[First]{Hugo O. Garcés} 
\author[Second]{Alejandro J. Rojas} 
\author[Third]{Bernardo A. Hernández Vicente}
\author[Third]{Andrés Escalona}
\author[Thirdddd]{Jonathan M. Palma}
\author[Fourth]{Md. Rezwan Parvez}
\author[Sixth]{Bhushan Gopaluni}
\author[Fifth]{Sirish L. Shah}

\address[First]{Departamento de Ingenier\'ia Inform\'atica y Ciencias de la Computaci\'on, Universidad de Concepci\'on, Concepci\'on, Chile (e-mail: hugarces@udec.cl)}
\address[Second]{Departmento de Ingenier\'ia El\'ectrica, Universidad de Concepci\'on, Concepci\'on, Chile (e-mail: arojasn@udec.cl)}
\address[Third]{Departamento de Ingenier\'ia Mec\'anica, Universidad de Concepci\'on, Concepci\'on, Chile (e-mail: \{behernandez,andrescalona\}@udec.cl)}
\address[Thirdddd]{Electrical Engineering Departament, Universidad de Talca, e-mail:  jonathan.palma@utalca.cl)}
\address[Fourth]{Department of Electrical \& Computer Engineering, University of Alberta, Edmonton, T6G 1H9, Alberta, AB, Canada (e-mail: mdrezwan@ualberta.ca)}
\address[Sixth]{Department of Chemical and Biological Engineering, University of British Columbia, Vancouver, BC V6T 1Z3, Canada (bhushan.gopaluni@ubc.ca)}
\address[Fifth]{Department of Chemical \& Materials Engineering, University of Alberta, Edmonton, T6G 1H9, Alberta, AB, Canada (e-mail: sirish.shah@ualberta.ca)}

\begin{abstract} 
This paper compares the performance of model-free controllers on a nonlinear system under cyberattacks, including false data injection and denial-of-service attacks. Four RL reward types are analyzed for accuracy, cost, and resilience. Results show that the Lyapunov reward offers the best resilience with low tracking error. Exponential mode also provides good trade-offs with acceptable resilience under moderate training conditions. Progressive and linear rewards converge faster but are less robust. RL-MPCs show strong steady-state resilience but require longer training times; RL-PID controllers are faster with significantly less training time. Proximal Policy Optimization outperforms Deep Deterministic Policy Gradient with a significant reduction in KPI variance. This study serves to highlight how well-designed RL rewards can improve performance and resilience against cyber threats

\end{abstract}


\end{frontmatter}

\section{Introduction}

Modern control applications increasingly involve nonlinear, uncertain, and time-varying dynamics operating in data-limited environments. Model-based strategies, although theoretically well-founded, often struggle when plant models are incomplete, computationally intensive, or subject to unmodeled disturbances. Model-free approaches based on Reinforcement Learning (RL) have recently emerged as an attractive alternative for achieving adaptive, optimal control directly from system data. In cyber–physical systems (CPS), these capabilities are essential, as communication delays, stochastic perturbations, and cyber threats continuously modify the effective dynamics. Consequently, designing resilient controllers that preserve stability, safety, and efficiency under evolving conditions is a critical challenge that requires bridging data-driven adaptation with classical control concepts to ensure robustness and autonomy in real-time operation.

RL can be interpreted as a unifying framework that generalizes classical control methodologies, since Linear Quadratic Regulator (LQR) and $\mathbb{H}_\infty$ control can be recovered as special cases when the system is linear, the dynamics are known, and the reward is quadratic. In contrast, when the model is unknown or partially observable, RL serves as a model-free optimal control scheme, learning the policy directly from closed-loop experience while preserving the recursive structure of dynamic programming \citep{delReal2022review,Zhu2025,yin2025,Li2023RLControl}. All these features make model-free RL control particularly attractive for systems operating on cyber–physical environments, where accurate modeling is complex and resilience to disturbances or attacks must be learned adaptively from data rather than derived analytically \citep{rieger2021}.

The main contribution of this work is to analyze the performance of model-free RL controllers in CPS under cyberattacks, while recognizing that no single method can optimize resilience, cost, and performance all at once. We propose a unified benchmarking framework that tests all controllers with the same disturbances, adversarial signals (cyberattacks), and model uncertainties. An additional contribution is the analysis of reward function design: instead of treating rewards just as performance indicators, we show that their structure strongly influences the learned policy, embedding both robustness and efficiency into the training process. By comparing different reward designs and analyzing their effects, we identify reward shaping as a crucial factor for the practical deployment of model-free RL in control tasks.

\section{Background}\label{sec:back}
We consider a discrete-time single-input single-output (SISO) dynamic system as in eq.(\ref{eq:plant}):
\begin{equation}
	x_{k+1} = f(x_k,u_k), 
	\qquad 
	y_k = h(x_k),
	\label{eq:plant}
\end{equation}
where $x_k \in \mathbb{R}^{n}$ represents the state of the controlled dynamic system, $u_k \in \mathbb{R}$ is the control input, and
$y_k \in \mathbb{R}$ is the measured output. The internal dynamics $f(\cdot)$ and $h(\cdot)$ are unknown and not explicitly identified~\citep{Precup2022DataDrivenMFC,KhakiSedigh2024DDC}.

\subsection{Model-free RL PID control systems}
At the core of the model-free RL PID controller, we retain a conventional discrete-time PID law, which acts as a robust backbone for tracking and disturbance rejection, such as described in eq.(\ref{eq:pid-law})
\begin{equation}
    u_k = K_{p,k} e_k + K_{i,k} I_k + K_{d,k} D_k,
    \label{eq:pid-law}
\end{equation}
where $e_k = r_k - y_k$ is the error between the external reference $r_k$ and the measured output $y_k$,  $I_k = \mathrm{sat}\big(I_{k-1} + e_k T_s\big)$ is the integrative term, $D_k = (1-\alpha_D) D_{k-1} + \alpha_D \frac{e_k - e_{k-1}}{T_s}$ is the derivative term, $T_s$ is the sampling time, $\alpha_D \in (0,1]$ is a derivative coefficient and $\mathrm{sat}(\cdot)$ denotes integral windup protection. In eq.(\ref{eq:pid-law}), the PID gains $K_{p,k}$, $K_{i,k}$ and $K_{d,k}$ are adaptive, where a reinforcement learning policy updates them at every step:
\begin{equation}
    K_{\bullet,k} = \Pi_{\mathcal{K}}\!\big( K_{\bullet,k-1} + a_{\bullet,k} \big),
    \qquad
    \bullet \in \{p,i,d\},
    \label{eq:gain-update}
\end{equation}
where $a_k = [a_{p,k},a_{i,k},a_{d,k}]^\top$ is the action calculated by the RL agent, and $\Pi_{\mathcal{K}}(\cdot)$ projects the gains onto admissible bounds  $\mathcal{K} = [K_p^{\min},K_p^{\max}] \times [K_i^{\min},K_i^{\max}]
\times [K_d^{\min},K_d^{\max}]$ to preserve closed-loop safety. From the PID structure, we obtain smooth tracking and robustness, while the RL agent learns how to reshape the gains over time to cope with operating-point changes, CPS-induced distortions, and cyber-physical attacks, without an explicit plant model \citep{KhakiSedigh2024DDC}. From the RL agent's perspective, it interacts with the closed-loop system through a PID observations vector $o_{PID~k}$ as defined in eq.(\ref{eq:observation1}-\ref{eq:observation2}):
\begin{equation}
    o_{PID~k} = 
    \begin{bmatrix}
        z_{PID~k} & K_{p,k} & K_{i,k} & K_{d,k} & s_k 
    \end{bmatrix}^\top
    \label{eq:observation1}
\end{equation}

\begin{equation}
    z_{PID~k} = \begin{bmatrix} y_k & r_k & e_k & u_{k-1} \end{bmatrix}^\top,
    \label{eq:observation2}
\end{equation}
where $z_{PID~k}$ is an online normalized vector of measured variables suitable for RL PID control, and $s_k \in [0,1]$ encodes the relative time within an episode of length $N$. The RL agent calculates a continuous action $a_k = \mu_\theta(o_k) + \varepsilon_k$ with policy parameters $\theta$ and exploration noise $\varepsilon_k$. In this article, two complimentary actor-critic algorithms are performed, PPO and DDPG. On the first hand, PPO is featured by a stochastic Gaussian policy $a_k \sim \mathcal{N}\!\big(\mu_\theta(o_k), \Sigma_\theta(o_k)\big)$ which is optimized via a clipped-surrogate objective that limits the Kullback--Leibler divergence between successive policies, which typically leads to stable learning in continuous control tasks~\citep{Li2023RLControl,Belousov2021RLAlgos}. On the second hand, DDPG is featured by a deterministic policy $a_k = \pi_\theta(o_k)$ where the exploration is controlled by an external noise process (e.g.\ Ornstein--Uhlenbeck), obtaining faster convergence and high-resolution tuning of the gains, at the expense of being more sensitive to reward design and noise~\citep{Li2023RLControl,LiQiu2019RLCPS}.

\subsection{Model-free RL MPC control systems}

The core of this method consist on instead of identifying a predictive model for control such as presented in eq.(\ref{eq:plant}), we combine a fixed baseline MPC formulation with a reinforcement learning (RL) agent that adjusts selected MPC tuning parameters online. This formulation provides a model-free architecture where prediction and optimization rely on a nominal model, but the closed-loop behaviour is steered by the RL policy, based exclusively on measured data and reward feedback. As in a MPC controller, eq.(\ref{eq:mpc-qps}) solves a quadratic program as follows:
\begin{equation}
	\begin{aligned}
		\min_{\{u_{k+i|k}\}} \quad &
		\sum_{i=0}^{N_y-1} 
		w_y \, (y_{k+i|k} - r_{k+i|k})^2
		+\\
        & \sum_{i=0}^{N_u-1} 
		w_{\Delta u} \, (u_{k+i|k} - u_{k+i-1|k})^2,
		\\
		\text{s.t.} \quad &
		x_{k+i+1|k} = A x_{k+i|k} + B u_{k+i|k}, \\
		&
		u_{\min} \le u_{k+i|k} \le u_{\max}, \\
		&
		x_{k|k} = x_k,
	\end{aligned}
	\label{eq:mpc-qps}
\end{equation}
where $(A,B)$ is a nominal linear model used solely inside the optimizer,
$N_y$ and $N_u$ are the prediction and control horizons, and $w_y$ and $w_{\Delta u}$
are the output-tracking and move-suppression weights. In MPC, only the first optimal input
$u_{k|k}^\star$ is applied to the plant. Although this structure resembles
standard predictive control, the key novelty lies in how $w_y$ and $w_{\Delta u}$
are adapted online with RL methods. Then, an RL agent updates the weight vector of MPC such as described in eq.(\ref{eq:mpc-weights})
\begin{equation}
	W_k = 
	\begin{bmatrix}
		w_{y,k} \\
		w_{\Delta u,k}
	\end{bmatrix},
	\label{eq:mpc-weights}
\end{equation}
based on real-time measurements. The policy receives an MPC observations vector $o_{MPC~k}$ which is composed by the terms in eq.(\ref{eq:observation_mpc1})
\begin{equation}
	o_{MPC~k} =  
	\begin{bmatrix}
		z_{MPC~k} & w_{y,k} & w_{\Delta u,k} & s_k
	\end{bmatrix}^\top
	\label{eq:observation_mpc1}
\end{equation}
	
	\begin{equation}
	z_{MPC~k} = \begin{bmatrix} y_k & r_k & e_k & u_{k-1} \end{bmatrix}^\top,
	\label{eq:observation_mpc2}
\end{equation}
where $z_{MPC~k}$ in eq.(\ref{eq:observation_mpc2}) encodes online-normalized variables suitable for MPC model-free control, and $s_k = k/N$ is the normalized episode time. The RL agent outputs a continuous action $a_k = [\Delta w_{y,k},\Delta w_{\Delta u,k}]^\top$, so that the updated weights are calculated as in eq.(\ref{eq:weight-update}):
\begin{equation}
	W_{k+1} = \Pi_{\mathcal{W}}\!\left( W_k + a_k \right),
	\label{eq:weight-update}
\end{equation}
Update weights $W_{k+1}$ from eq.(\ref{eq:weight-update}) are projected into an admissible region $\mathcal{W}$ to prevent the optimization from becoming too aggressive or ill-conditioned. The above means that the RL agent does not replace the optimizer; rather, learns how the MPC should behave to ensur stability and robustness against distortions, becoming as a way of supervisor that adapts tracking aggressiveness, disturbance rejection and control smoothness without an explicit system identification phase. In this article, two complementary algorithms are performed for the model-free MPC controller, PPO, and DDPG. On the first hand, PPO uses a Gaussian policy $a_k \sim \mathcal{N}\big(\mu_\theta(o_k),\Sigma_\theta(o_k)\big)$ optimized via a clipped-surrogate objective. This tends to produce stable, conservative adaptations of $W_k$, especially under noisy CPS channels or latency variations, consistent with the robustness properties reported in continuous-control literature~\citep{Li2023RLControl,Belousov2021RLAlgos}. On the second hand, DDPG uses a deterministic policy $a_k = \pi_\theta(o_k)$ complemented by an exploration noise process. DDPG and its deterministic  policy yields faster adaptations in the model-free RL MPC controller, enabling to sharpen transients or reduce overshoot when the reward encourages more aggressive behaviour~\citep{LiQiu2019RLCPS}. 


\subsection{Reward function design and closed-loop behaviour}

For model-free RL PID and MPC control systems, the final critical aspect for design is the reward function, where $k-th$ time defines a scalar reward signal such as defined in eq.(\ref{eq:reward-generic1}): 

\begin{equation}
	r_k = r_k^{\mathrm{perf}}(e_k) 
	- \beta_{\Delta u} \big|u_k - u_{k-1}\big|
	- \gamma_u u_k^2,
	\label{eq:reward-generic1}
\end{equation}
where hyperparameters $\beta_{\Delta u}$ and $\gamma_u$ are tuned to balance transient speed against smooth and energy-efficient response of the controller. Here, the first term of the reward function $r_k^{\mathrm{perf}}(e_k)$ is designed according to a set of expressions focused on robustness, speed, and accuracy. In all cases, reward function $r_k$ in eq.(\ref{eq:reward-generic1}) captures tracking performance, while the two regularizers penalize aggressive actuation and excessive controller effort. Conceptually, this reward acts as a surrogate optimality criterion that guides the RL agent to reshape the PID or MPC hyperparameters. From a control point of view, reward function design can be seen as data-driven surrogates for traditional quadratic costs, now expressed directly in the framework of RL~\citep{Belousov2021RLAlgos,Lian2024IntegralInverseRL}. In practice, RL-based tuning allows the PID controller to adapt its gain schedules to reduce error while balancing overshoot and settling time against control smoothness. For RL MPC, the RL-driven structure makes it possible to adapt across operating regimes—acting more aggressively when rapid tracking is safe, and more conservatively when noise, delays, or adversarial inputs prevail. This highlights a key principle of the RL–MPC framework: MPC furnishes the optimization machinery, whereas RL provides the closed-loop intelligence to reshape the cost function online using only data and general-purpose learning methods.

\section{Methodology}\label{sec:method}
\subsection{Model-free control boosted by reinforcement learning}

Model-free control boosted by reinforcement learning provides a bridge between adaptive control theory and modern data-driven optimization. From a control standpoint, RL approximate value functions or policies are iteratively updated using data collected from closed-loop operation, which represents the environment where RL learns. Here, the control policy is optimized directly from interaction with the environment, without an explicit mathematical model of the plant. The objective is defined by a cumulative reward (or cost) that encodes tracking accuracy, control effort, and robustness to disturbances, and the controller seeks a policy \(\pi(a|s)\) that maximizes
\begin{equation}
J(\pi) = \mathbb{E}\left[\sum_{t=0}^{\infty} \gamma^t R(s_k, a_k)\right].
\label{eq:rl_objective}
\end{equation}
Unlike model-based approaches such as MPC, which solve an optimization problem using a predictive model, model-free RL updates the policy parameters from data by approximating the Bellman equation and the gradient of \(J(\pi)\) in \eqref{eq:rl_objective}. Its main advantage is that it does not require an explicit plant model, allowing adaptation to nonlinear, time-varying processes with partially unknown dynamics \citep{lawrence2024, mcclement2022}. Once trained, the RL agent maps states directly to control actions with negligible online computational cost \citep{spielberg2019}, and systematic exploration–exploitation enables the controller to preserve performance.

\subsection{Reward functions for testing of model-free RL based controllers}

A modular reward design is adopted that separates a performance term from actuation regularizers, as previously defined in eq.(\ref{eq:reward-generic1}). The reward mode determines the shaping of the performance term $r_k^{\mathrm{perf}}(e_k)$ and represents a key factor on the operation of the model free controllers based on RL, since it determines if the controller is mainly designed for resilience and robustness, or for improved performance, where all modes share the same regularization terms and ensure comparable energy constraints across experiments \citep{Ibrahim2024,guerraoui2024robust,hrgovic2025reward}.

\subsubsection{Exponential (\texttt{exp})}
This mode applies an exponential penalty on the absolute error,
\begin{equation}
r_k^{\mathrm{perf}}(e_k) = -\big(\exp(\alpha\,|e_k|) - 1\big),
\end{equation}
where \(\alpha>0\) determines the curvature of the penalty.
Large deviations are sharply penalized, whereas small deviations remain sensitive.
This mode is particularly effective under high dynamic ranges or communication delays.

\subsubsection{Progressive (\texttt{pro})}
This baseline formulation rewards the progressive reduction of squared error between successive steps:
\begin{equation}
r_k^{\mathrm{perf}}(e_k)= \kappa\,(e_{k-1}^2 - e_k^2),
\end{equation}
where \(\kappa>0\) controls the reward gain. Positive reward is obtained only when the error decreases, which promotes monotonic convergence without explicit reference shaping.

\subsubsection{Lyapunov-descent (\texttt{lya})}
This design integrates a Lyapunov-inspired metric that combines instantaneous and integral error components:
\begin{equation}
V_k = e_k^2 + \lambda_I\,I_k^2, \qquad 
r_k^{\mathrm{perf}}(e_k,I_k) = \kappa\,(V_{k-1} - V_k),
\end{equation}
with \(\lambda_I>0\) weighting the integral term \(I_k\), and tuning gains \(\kappa\) and \(\lambda_I\). The reward is positive only when \(V_k\) decreases, which aligns the learning signal with classical stability conditions and avoids oscillatory behavior.

\subsubsection{Linear (\texttt{lin})}
Simple linear penalty is applied on the absolute error:
\begin{equation}
r_k^{\mathrm{perf}}(e_k) = -\alpha_{\mathrm{err}}\,|e_k|,
\end{equation}
with slope \(\alpha_{\mathrm{err}}>0\).


At the end of each episode, a terminal reward is added to encode overall success:
\begin{equation}
\begin{split}
R_T &= \mathrm{sign}(V_{\mathrm{ter}})\,|V_{\mathrm{ter}}|^{P_{\mathrm{obs}}},
\\
P_{\mathrm{obs}} &= \max\!\left\{0,\,1 - \min\!\big(1, |e_T|/e_{\max}\big)\right\},
\end{split}
\end{equation}
where \(R_{\mathrm{win}}>0\) and \(R_{\mathrm{fail}}<0\) define the bounds, and \(e_{\max}\) normalizes the proximity score. This term assigns smoothly interpolated credit or penalty based on the final tracking accuracy.

\subsection{Key Performance Indicators (KPIs) and benchmark controllers }

To evaluate the proposed model-free RL controllers, we benchmark them against three established control strategies: adaptive control, MPC, and PID regulation. Adaptive controllers update their parameters online through deterministic adaptation laws, yielding smooth compensation for plant changes but relying on structural assumptions and sufficient excitation. MPC constitutes a more powerful model-based reference, solving a finite-horizon optimal control problem at each sampling instant while enforcing dynamic and input constraints, and thus achieving high performance when accurate models and adequate computation are available. PID control serves as a simple, widely used baseline that requires no model knowledge and delivers reliable feedback under nominal conditions. The assessment of both model-free RL and benchmark controllers relies on Key Performance Indicators (KPIs) that reflect not only tracking accuracy but also computational efficiency and robustness to disturbances or cyberattacks. Accordingly, we group the KPIs into three categories: \textit{Error-based}, \textit{Computational-cost}, and \textit{Resilience-based} indicators, each comprising three analytically defined metrics that quantify desirable controller properties.

\subsubsection{Error-based KPIs}

Error-based indicators quantify how accurately the controlled output \(y_k\) follows the reference signal \(y_{ref~k}\). Let \(e_k = y_{ref~k} - y_k\) denote the instantaneous tracking error, and \(T_s\) the sampling time. Mean Squared Error (MSE) is defined as follows:
\begin{equation}
\label{eq:kpi_mse}
\mathrm{MSE} = \frac{1}{N}\sum_{t=1}^{N} e_k^2.
\end{equation}
Low MSE value indicates small average deviations and stable steady-state tracking. However, this index penalizes large errors quadratically, which can overweight outliers if the system experiences transient spikes. Next, Integral Absolute Error ($IAE$) represents the accumulated tracking error over the entire testing episode, which is defined as follows:

\begin{equation}
\label{eq:kpi_iae}
\mathrm{IAE} = \sum_{t=1}^{N} |e_k|\,T_s.
\end{equation}

Lower values correspond to faster settling and reduced overall deviation from the reference trajectory. $IAE$ is particularly useful when transient duration and steady-state accuracy are equally important.

Finally, controller effort ($\overline{\| \Delta U \|}$) quantifies the variability or aggressiveness of the control signal sequence, defined as follows:
\begin{equation}
    \overline{\| \Delta U \|} = \frac{1}{N-1} \sum_{k=1}^{N-1} \left( u_{k} - u_{k-1} \right)^2
\end{equation}

$\overline{\| \Delta U \|}$ quantifies the variation of the control action between consecutive sampling instants and indicates actuator stress and control smoothness, where $u_k$ is the control input at discrete time step $k$-th, and $N$ is the total number of control samples.

\subsubsection{Computational-cost KPIs}

These indicators evaluate the computational efficiency and feasibility of the control algorithm, measuring the numerical demand, real-time responsiveness, and hardware utilization, which are critical for cyber–physical systems operating under strict temporal constraints. First, the training time ($Time_{\mathrm{train}}$) measures the total duration required to complete the learning or optimization process of the control policy:
\begin{equation}
\label{eq:kpi_trainingtime}
Time_{\mathrm{train}} = t_{\mathrm{end}} - t_{\mathrm{start}},
\end{equation}
where \(t_{\mathrm{start}}\) and \(t_{\mathrm{end}}\) correspond to the initial and final timestamps of the training phase. Lower values of \(T_{\mathrm{train}}\) indicate faster convergence and improved algorithmic efficiency, whereas excessively high training durations may reveal suboptimal hyperparameter tuning or high model complexity. Secondly, the average CPU utilization (\(\overline{CPU}\)) quantifies the mean percentage of processor resources used throughout the experiment or real-time operation:
\begin{equation}
\label{eq:kpi_avgcpu}
\overline{CPU} = 
\frac{1}{N} \sum_{k=1}^{N} \Psi_{\mathrm{CPU}}(k),
\end{equation}
where \(\Psi_{\mathrm{CPU}}(k)\) represents the instantaneous CPU usage at step \(k\). High average CPU utilization may indicate heavy computational load or inefficient resource management, while moderate utilization levels (\(35\text{–}70\%\)) typically reflect a balanced compromise between responsiveness and system overhead. The maximum step time ($\Omega_{\mathrm{max}}$) captures the worst-case computational delay across all control iterations:
\begin{equation}
\label{eq:kpi_maxsteptime}
\Omega_{\mathrm{max}} = \max_{1 \le k \le N} \Omega_k.
\end{equation}

The combined analysis of these KPIs ensures that the learning-based controller is computationally viable for real-time cyber–physical systems or edge-AI controllers.

\subsubsection{Resilience-Based KPIs}

Resilience is a novel concept from cyber–physical system (CPS), which refers by its capacity to initially withstand or minimize the impact of a disturbance (resistance phase) and subsequently recover its performance, both in the short term or response phase, and the long term or restoration phase \citep{Vamvoudakis2021HandbookRLControl,BruntonKutz2022}. Firstly, we present the adaptive capacity $\mathrm{AdC}$ defined as follows:
\begin{equation}
    \mathrm{AdC}
    = \min\!\left\{1,\;
        \max\!\left\{0,\;
            1 - \frac{\mathrm{MSE}_{\text{post}}}{\max\bigl(\varepsilon_k,\;\mathrm{MSE}_{\text{pre}}\bigr)}
        \right\}
    \right\},
    \label{eq:adaptive_capacity}
\end{equation}
where $\mathrm{MSE}_{\text{pre}}$ is the mean squared tracking error computed over a time window preceding the attack, $\mathrm{MSE}_{\text{post}}$ is the mean squared tracking error over a post-recovery window, and $\varepsilon_k$ is a small positive constant used to avoid division by zero.  The outer $\min\{\cdot\}$ and $\max\{\cdot\}$ operators bound the index in the interval $[0,1]$.  When $\mathrm{MSE}_{\text{pre}} > 0$ the expression in \eqref{eq:adaptive_capacity} simplifies to a normalized performance difference between the pre-attack and post-recovery regimes. When $\mathrm{AdC} \approx 0$, the post–attack error satisfies 
$\mathrm{MSE}_{\text{post}} \gtrsim \mathrm{MSE}_{\text{pre}}$, which indicates poor adaptive capacity and shows that the controller cannot restore nominal performance after the attack. Conversely, when 
$\mathrm{AdC} \approx 1$ and $\mathrm{MSE}_{\text{post}} \ll \mathrm{MSE}_{\text{pre}}$, the CPS recovers effectively and the reconfiguration compensates the disturbance. For intermediate values $0 < \mathrm{AdC} < 1$, the system shows partial recovery, with post–attack error still larger than in the pre–attack condition.

Margin to maneuver ($MaM$) quantifies the remaining operational headroom before the system violates performance or safety constraints:
\begin{equation}
\label{eq:kpi_margin_maneuver}
MaM = 
\frac{P_{\mathrm{max}} - P(t)}{P_{\mathrm{max}} - P_{\mathrm{min}}},
\end{equation}
where \( P_{\mathrm{max}} \) and \( P_{\mathrm{min}} \) define the physical or operational limits of the system variable. A high margin (\( M_{\mathrm{man}} \to 1 \)) implies that the system operates well within safe bounds and retains flexibility to absorb further perturbations. On the other hand, when \( M_{\mathrm{man}} \to 0 \), the system has exhausted its controllable range and is at high risk of instability or mission failure. $AdC$ and $MaM$ characterize the dynamic resilience profile of a control system, where $AdC$ evaluates the ability of the controller to autonomously adapt, particularly relevant for RL-based and adaptive control policies, and $MaM$ represents the instantaneous safety margin, which reflects how close the system operates to performance or physical limits. 

\subsection{LPV Benchmark and CPS Attack Modeling}

The LPV benchmark used in our experiments corresponds to a first–order SISO plant whose dynamics vary through the effective time constant $\tau*$ where $\tau*=\bar{\tau}\pm\sigma_{\tau}$ and sampling period $T_s$. The discrete–time model is defined as eq(\ref{eq:LPVsys}):
\begin{equation}
x_{k+1} = \Bigl(1 - \tfrac{T_s}{\tau*}\Bigr)x_k + \tfrac{T_s K}{\tau*}u_k
\label{eq:LPVsys}
\end{equation}
with $K$ the static gain and the output $y_k=x_k$. Besides, a realistic CPS communication channel is explicitly simulated at every sampling
instant, where the true output is corrupted by Gaussian noise, followed by packet dropout that holds the last received value. A random network delay in $[0,\,2T_s]$ is imposed through a FIFO buffer, and a quasi–static drift is accumulated after a prescribed time step to emulate long–term sensor bias. In the closed loop, additional cyber–physical
attacks may occur: (i) drift attacks: introduce an ever–increasing bias;
(ii) noise attacks: inject amplified Gaussian disturbances; and (iii) DoS
attacks: freeze the transmitted measurement (hold–last–sample). This combination of noise, dropout, delay, and biased feedback provides a compact but challenging setting to evaluate the resilience of model–free RL controllers.

\section{Results}\label{sec:results}
\subsection{RL-PID}

\begin{figure}
    \centering
    \includegraphics[width=0.99\linewidth]{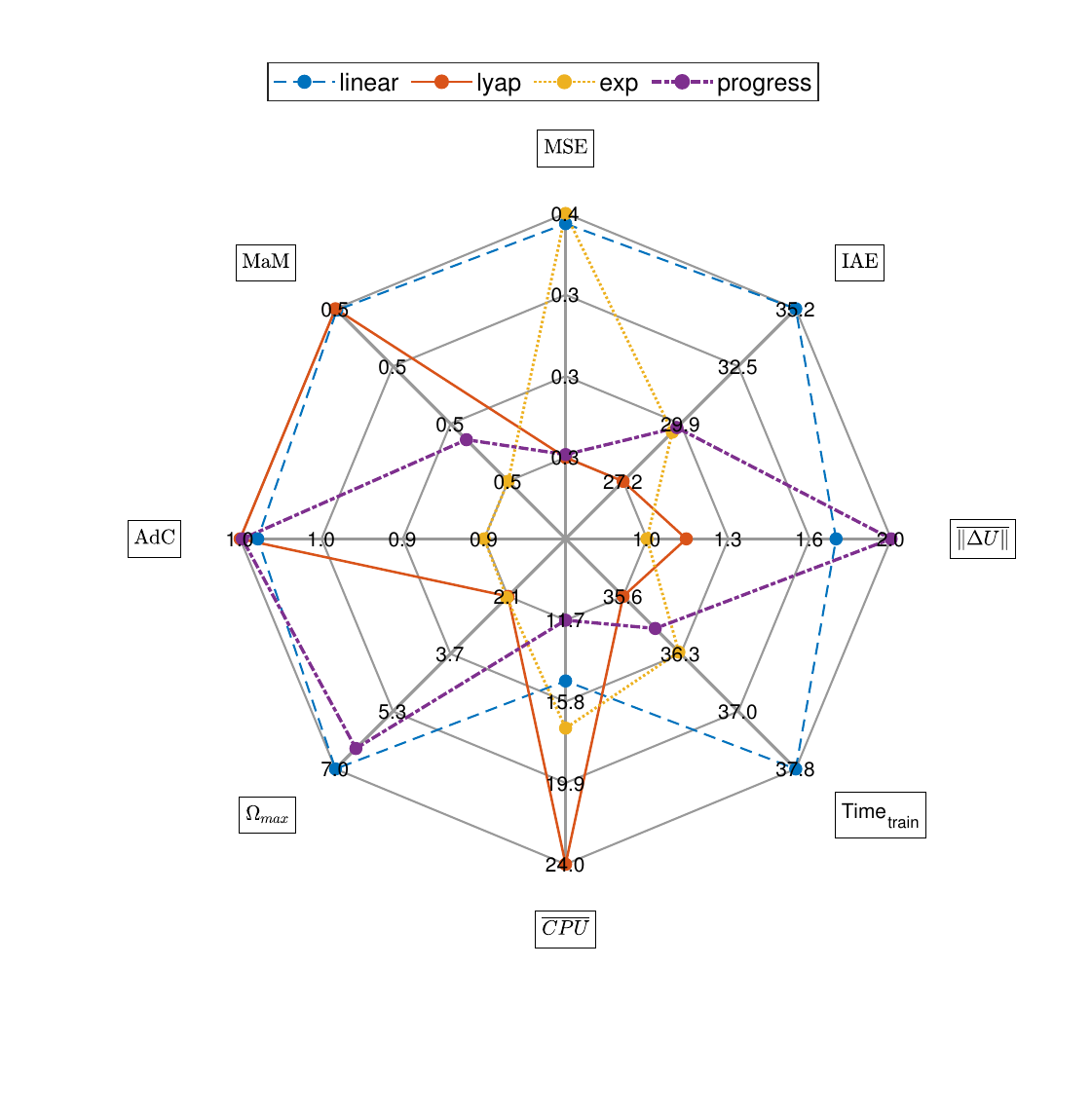}
    \caption{KPI's summary RL-PID controller trained with PPO }
    \label{fig:rl_pid_ppo}
\end{figure}

\begin{figure}
    \centering
    \includegraphics[width=0.99\linewidth]{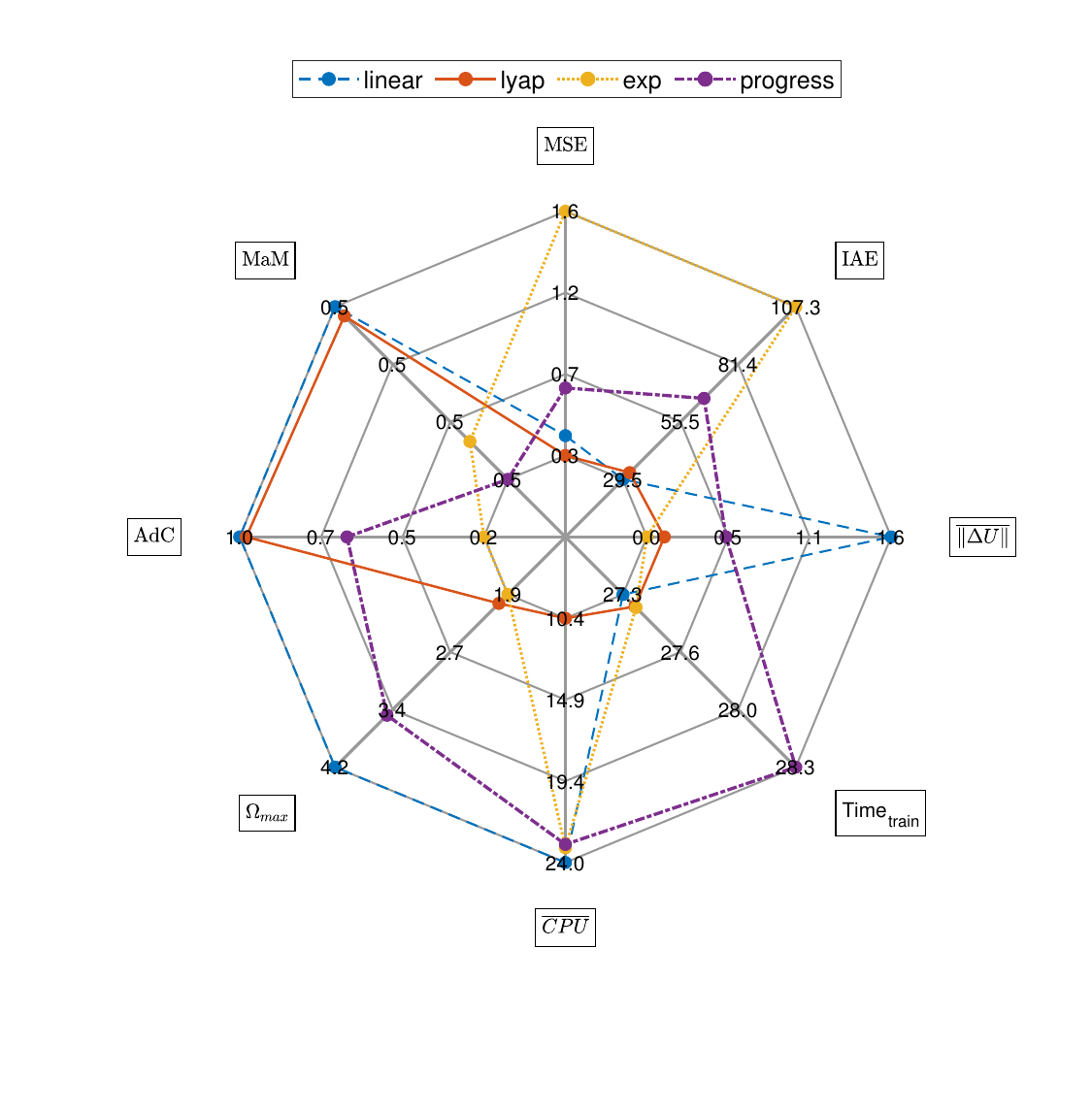}
    \caption{KPI's  summary RL-PID controller trained with DDPG }
    \label{fig:rl_pid_ddpg}
\end{figure}

Figure \ref{fig:rl_pid_ppo} presents the radar plot of the key performance indicators (KPIs) for the RL-PID controller trained with the PPO algorithm, highlighting the balance achieved by the Lyapunov and exponential modes, which maintain moderate error levels while preserving robustness and acceptable recovery dynamics. Figure \ref{fig:rl_pid_ddpg} summarizes the performance of the RL-PID controller when trained with DDPG. Compared to PPO, DDPG exhibits higher variability across reward modes, especially in training time and recovery duration. Lyapunov-based reward yields the most stable control responses across all metrics.

\subsection{RL-MPC}

\begin{figure}
    \centering
    \includegraphics[width=0.99\linewidth]{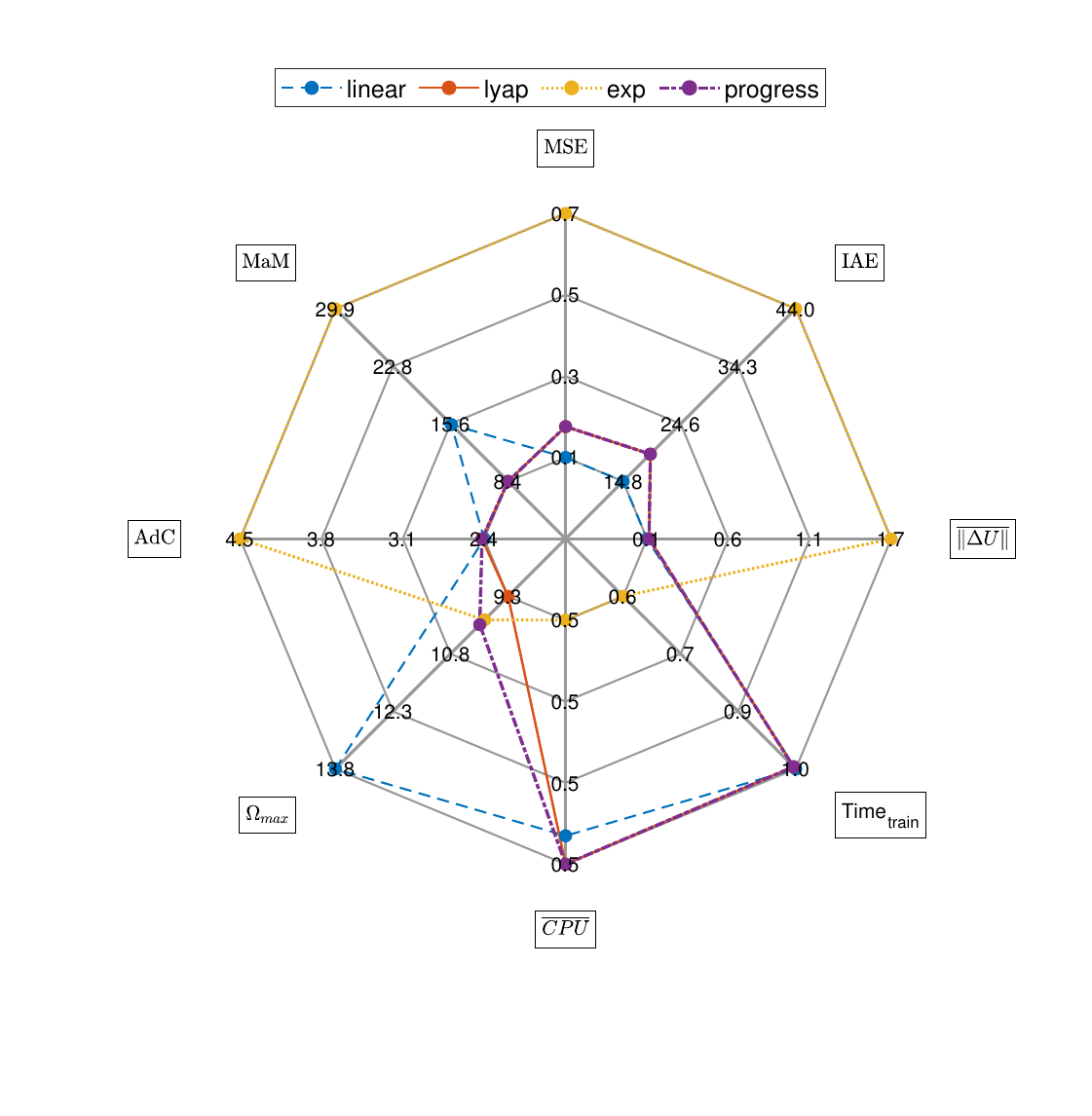}
    \caption{KPI's summary RL-MPC controller trained with PPO }
    \label{fig:rl_mpc_ppo}
\end{figure}

\begin{figure}
    \centering
    \includegraphics[width=0.99\linewidth]{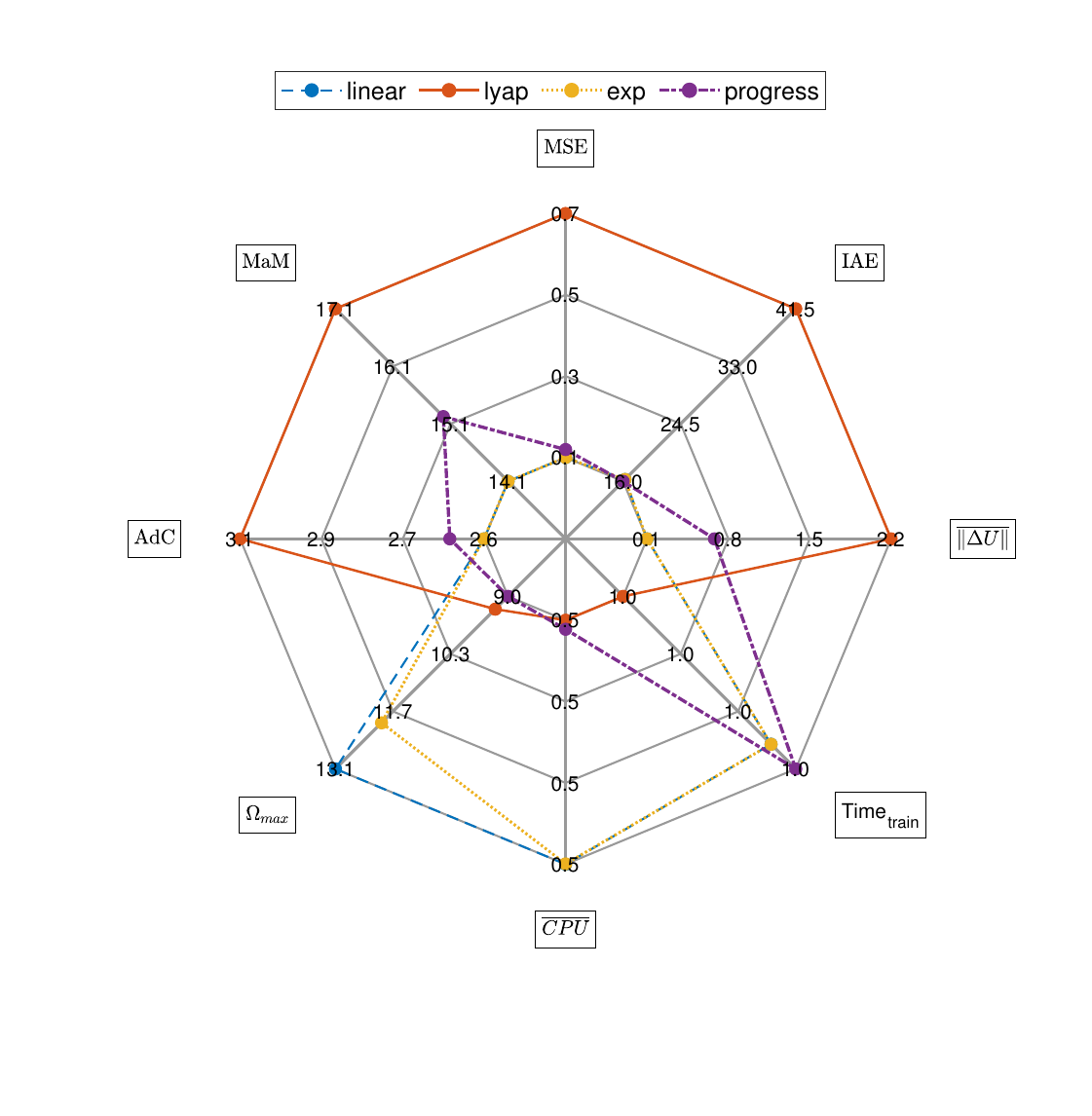}
    \caption{KPI's summary RL-MPC controller trained with DDPG }
    \label{fig:rl_mpc_ddpg}
\end{figure}

Figure \ref{fig:rl_mpc_ppo} illustrates the KPI distribution for the RL-MPC controller trained with PPO. The radar plot shows improved overall consistency compared to RL-PID, particularly in the ReI and MaM indicators. The exponential and balanced modes deliver an optimal compromise between fast recovery and computational efficiency. Figure \ref{fig:rl_mpc_ddpg} depicts the RL-MPC controller trained with DDPG shows the widest performance dispersion among the four tested combinations and reflects the sensitivity of DDPG to reward shaping. Although the progressive and balanced modes promote rapid convergence, their resilience under attack is notably lower. In contrast, the Lyapunov and exponential modes sustain better stability and recovery at the expense of longer training times.

\subsection{Benchmarking results}
Based on the analyzed results, two controllers were selected to highlight the contrast between resilience-oriented and error-oriented RL model-free control systems. For the resilience benchmark, the RL-MPC control system trained with PPO using the \texttt{lyap} reward is selected, as it consistently achieves the highest ReI and MaM values and exhibits the most stable radar profile across all test cases. As a counterpart, the error-focused benchmark is represented by the RL-PID controller trained with DDPG under the \texttt{progress} reward.
This configuration delivers fast convergence and low tracking error, yet it shows reduced robustness under disturbances and attacks. The selected controllers provide a basis for comparative analysis of stability-driven and
performance-driven RL control strategies.

\begin{figure}
   \centering
   \includegraphics[width=0.99\linewidth]{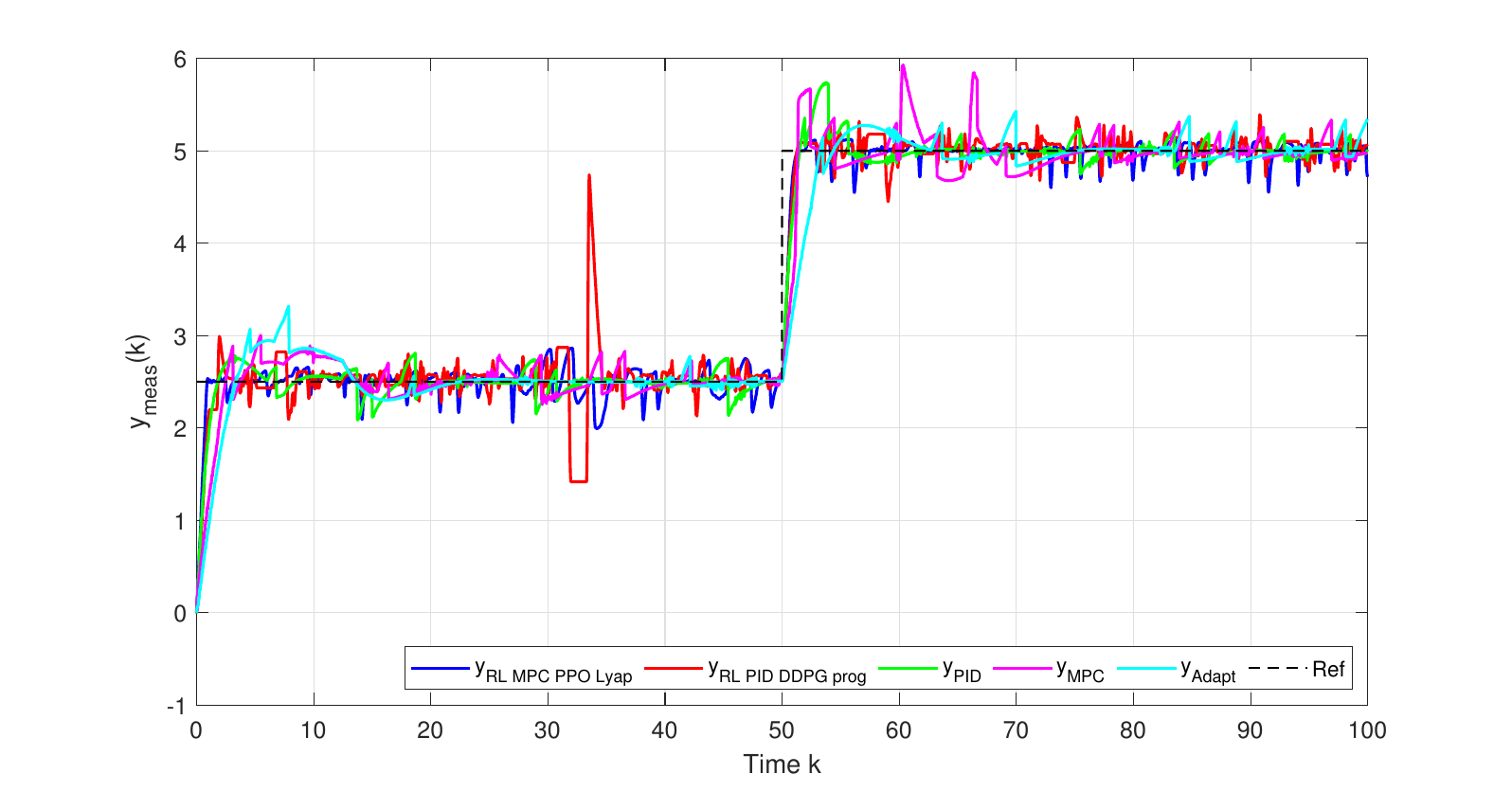}
   \caption{Dynamic behavior of RL model-free and reference controllers}
   \label{fig:dynamic_response}
\end{figure}

\begin{figure}
   \centering
   \includegraphics[width=0.99\linewidth]{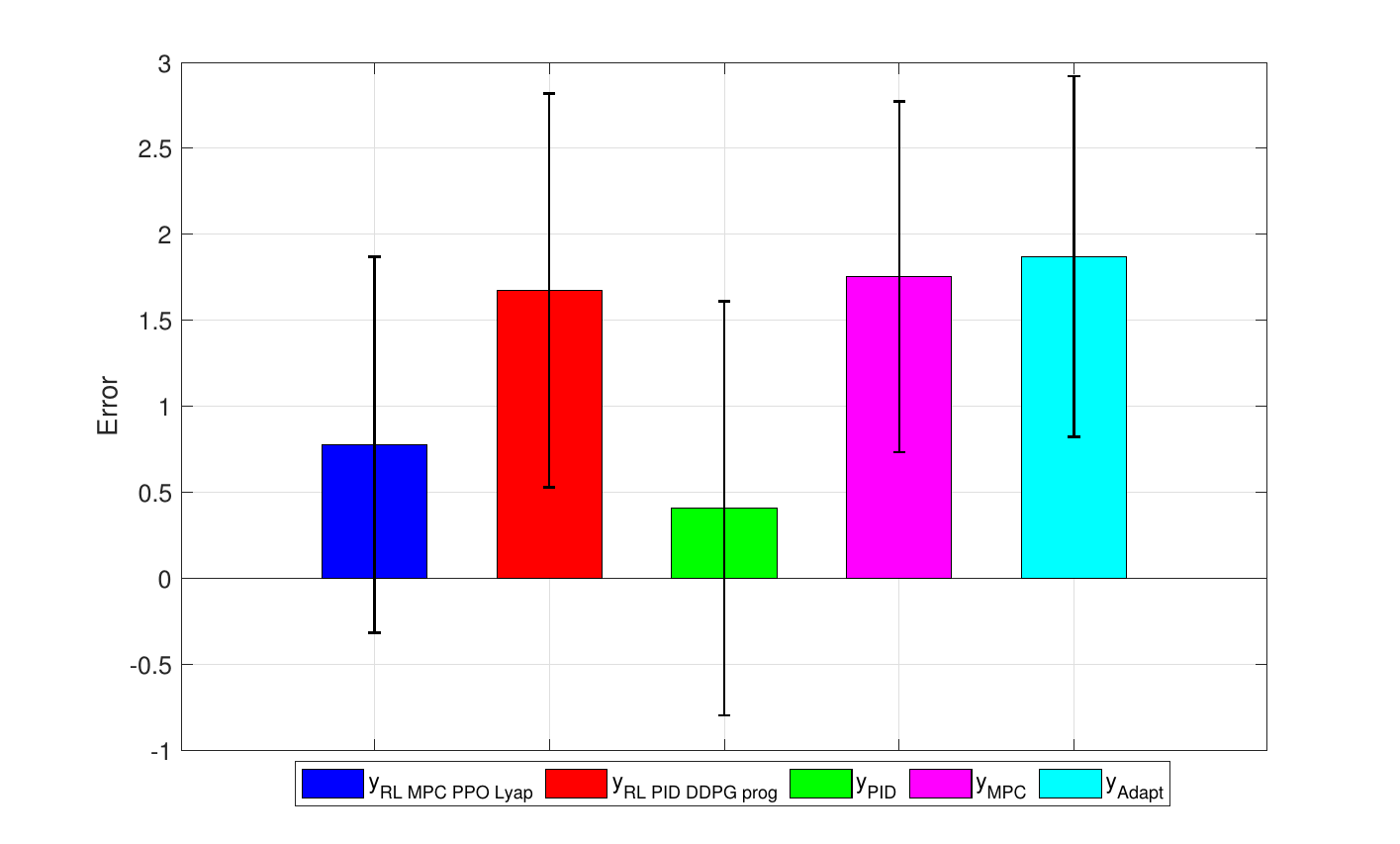}
   \caption{Mean error and standard deviation of model-free RL and benchmark controllers}
   \label{fig:error}
\end{figure}

\section{Analysis \& Conclusions}

The radar plots in Figures~\ref{fig:rl_pid_ppo}–\ref{fig:rl_mpc_ddpg} reveal clear trends in how each reward function shapes controller behaviour. The \texttt{lyap} mode consistently yields the most stable and resilient responses, with large $AdC$ and $MaM$ values and low MSE and IAE, indicating smooth transients and reliable post-attack recovery. The \texttt{exp} mode follows closely, offering a balanced compromise between accuracy and robustness. In contrast, \texttt{progress} emphasizes fast convergence but exhibits reduced resilience under noise or delay, especially with DDPG. The linear baseline remains the least adaptive, with higher steady-state error and slower recovery. Across learning methods, PPO produces smoother and more coherent KPI profiles, while DDPG shows higher variability and stronger sensitivity to reward curvature. Structurally, RL-MPC offers higher resilience, whereas RL-PID remains computationally lighter and well suited for embedded applications. Overall, the reward hierarchy observed is:
$\texttt{lyap} \rightarrow \texttt{exp} \rightarrow \texttt{progress} \rightarrow \texttt{linear}$ capturing the trade-off between robustness and aggressiveness and reinforcing the importance of reward design in RL-based CPS control. In addition, Figures \ref{fig:dynamic_response}-\ref{fig:error} further illustrate the differences among controllers, with RL-MPC showing the smoothest transients and smallest deviations during disturbances. RL-PID remains accurate but responds more sharply around attack intervals, while adaptive and Koopman controllers exhibit larger excursions, especially under drift and DoS. Besides, RL-MPC and RL-PID achieve the lowest mean error and tightest variability, confirming their strong noise rejection. This study confirms that no controller simultaneously achieves maximum resilience, low computational cost, and minimal tracking error; each approach excels under specific operating conditions. Reward shaping emerges as a central mechanism for enhancing model-free RL performance, enabling controllers to encode objectives—such as resilience to noise, drift, and DoS attacks—that are difficult to express with traditional feedback laws. Future work will explore additional reward formulations and pursue analytical stability guarantees for model-free RL control.

\section{Acknowledgments}
The authors acknowledge the financial support of ANID (Chile) through grants Fondecyt Regular 1220903 (Hugo O. Garc\'es), 1230777 (Alejandro J. Rojas), 1241305 \newline (Jonathan M. Palma); and Doctoral Thesis in the Productive Sector 2024 88240006 (Andr\'es Escalona).

\bibliography{REFS}

\end{document}